\documentclass[aps,floatfix]{revtex4}

\usepackage{amssymb,hyperref,hypcap,
ae,tikz,color}
\usepackage{latexsym,textcomp,slashed,epsfig}
\usepackage[T1]{fontenc}
\usepackage{verbatim}

\usepackage{dcolumn}
\usepackage{bm}
\usepackage{amsmath}
\usepackage{graphicx}
\usepackage{listings}
\usepackage[T1]{fontenc}
\usepackage{xcolor}
\usepackage{lmodern}
\usepackage{listings}

\hypersetup{
	colorlinks=true,
	citecolor=green,
    linkcolor=magenta,
	}

\begin{document}

\title{{\bf Axial Gravitational Waves in Bianchi I Universe}}

\author{\bf{Sarbari Guha and Sucheta Datta}}

\affiliation{\bf Department of Physics, St.Xavier's College (Autonomous), Kolkata 700016, India}

\maketitle

\section*{Abstract}
In this paper, we have studied the propagation of axial gravitational waves in Bianchi I universe using the Regge-Wheeler gauge. In this gauge, there are only two non-zero components of $ h_{\mu\nu} $ in the case of axial waves: $h_0(t,r)$ and $h_1(t,r)$. The field equations in absence of matter have been derived both for the unperturbed as well as axially perturbed metric. These field equations are solved simultaneously by assuming the expansion scalar $\Theta$ to be proportional to the shear scalar $\sigma$ (so that $a= b^n$, where $a$, $b$ are the metric coefficients and $n$ is an arbitrary constant), and the wave equation for the perturbation parameter $h_0(t,r)$ have been derived. We used the method of separation of variables to solve for this parameter, and have subsequently determined $h_1(t,r)$. We then discuss a few special cases in order to interpret the results. We find that the anisotropy of the background spacetime is responsible for the damping of the gravitational waves as they propagate through this spacetime. The perturbations depend on the values of the angular momentum $l$. The field equations in the presence of matter reveal that the axially perturbed spacetime leads to perturbations only in the azimuthal velocity of the fluid leaving the matter field undisturbed.

\bigskip

KEYWORDS: General Relativity; Bianchi I spacetime; Axial gravitational waves; Regge-Wheeler gauge

\section{Introduction}

The theory of gravitational waves (GWs), originally proposed by Einstein, was based on a linearization of the gravitational field equations \cite{EINST1,EINST2}. But Einstein was doubtful whether the fully nonlinear field equations admit solutions which represent GWs. The first attempt to define a plane gravitational wave in the full nonlinear theory was made by Rosen \cite{Rosen} and by Einstein and Rosen \cite{ER} in 1937. In order to study the dynamics of the gravitational field, and the corresponding stability problems, it was necessary to formulate and solve the Cauchy problem on the evolution of gravitational dynamics from an initial data. In 1952 Yvonne Choquet-Bruhat proved the existence of solutions to the nonlinear Einstein equations (EFEs) for a given Cauchy data \cite{Yvonne}. In 1957 Bondi \cite{Bondi} showed that GWs, which carry energy, do actually exist. In the same year, Pirani \cite{Pirani} offered an invariant formulation of gravitational radiation. Together with Pirani and Robinson, Bondi demonstrated that plane GWs are non-flat solutions of Einstein's vacuum field equations and possess the same symmetry as electromagnetic waves \cite{BPR}. Subsequently, Robinson and Trautman derived a class of solutions to Einstein's equations in vacuum, which represented spherical waves \cite{RT1}. They also solved the EFEs for the class of metric which could be considered to represent simple spherical outgoing waves \cite{RT2}.

The formulation of GWs become complicated due to the nonlinearity of the EFEs. To address this problem, one has to adopt approximation methods \cite{ST}. The theory of black hole perturbations was originally developed as a metric perturbation theory. Regge and Wheeler used this method to examine the stability of Schwarzschild singularity \cite{RW}. They introduced small perturbations into the background expressed in terms of spherical harmonics, and decomposed them into tensor harmonics. This enabled them to derive a single Schr\"{o}dinger-type differential equation for the perturbations. The perturbation equations were written in standard Schwarzschild coordinates $(t, r, \theta, \phi)$, using a standard gauge (known as the ``Regge-Wheeler gauge''). Thus they could extract two gauge-independent linearized modes, the `axial' and the `polar' mode. Zerilli \cite{ZER} investigated the problem of a particle falling into a Schwarzschild black hole using this method and also corrected the propagation equation for polar waves derived in \cite{RW}.

Vishveshwara \cite{VISH} followed the Regge-Wheeler (RW) procedure in Kruskal coordinates to examine the stability of the Schwarzschild exterior metric against small perturbations. The metric perturbation theory has been summarily presented by Chandrasekhar \cite{Chandra}.
Martel and Poisson presented a covariant and gauge-invariant formalism to study the metric perturbations of the Schwarzschild spacetime \cite{MP}.
Viaggiu \cite{VIA} used Laplace transforms to derive the RW equation for axial and polar waves in a de Sitter universe. The Laplace-transformed axial and polar perturbations were found to be functions of retarded and advanced metric coefficients with respect to the Laplace parameter $s$. The RW and Zerilli equations have been re-derived using variational principle by Moncrief \cite{MON}. Fiziev \cite{FIZ} determined the exact solutions of the RW equation describing the axial perturbations of the Schwarzschild metric in linear approximation in terms of the confluent Heun's functions, HeunC($ \alpha, \beta, \gamma, \delta, \eta, r$).

In the cosmological context, Malec and Wylezek \cite{M1} used the RW scheme to derive a propagation equation for axial modes in the FLRW spacetime with vanishing cosmological constant. The GWs obeyed Huygens principle in the radiation-dominated era, but in the matter-dominated universe, the propagation depends on their wavelengths. Short waves practically satisfy Huygens principle whereas long waves can backscatter off the curvature of spacetime. The backscattered axial waves becomes significant only for wavelengths comparable to the Hubble radius.

Kulczycki and Malec extended the work of \cite{M1} to study polar GWs in FLRW spacetimes \cite{M2}. The polar waves perturb both the density and non-azimuthal components of the velocity of the material medium, leading to the evolution of matter inhomogeneities and anisotropies. However, in the case of axial waves, the initial data could be chosen such that they decouple from matter, influencing only the azimuthal velocity and cause local cosmological rotation. The Huygens principle was valid for both waves only in the radiative spacetimes with vanishing cosmological constant.
It was shown in \cite{M3} that smooth axial GWs do not interact with perfect fluids in radiative FLRW spacetimes, and does not affect the relative distance between two test bodies. But if the initial profiles are not smooth, then the wave pulses force the radiation fluid to rotate, and the rotation persisted after the pulse moved away. Sharif and Siddiqa discussed the propagation of axial waves in \cite{SH1} and polar waves in \cite{SH2} using the RW gauge in FLRW spacetime in the context of $ f(R,T) $ gravity. Their study shows that the properties of the two types of waves match with those deduced in \cite{M2,M3} in the context of general relativity.

Axial and polar perturbations have also been studied in gauge-invariant formalism \cite{GMG, GS, CL1}. Clarkson et al. \cite{CL1} presented a full system of master equations that represent the general perturbations to dust dominated LTB space-times. Any perturbation in FLRW cosmology can be split into scalar, vector and tensor (SVT) modes that decouple from each other and evolve independently. A similar split could not be performed in the LTB model as the background is not spatially homogeneous and the modes get coupled. However, depending on the nature of transformation of the perturbations on the surfaces of spherical symmetry, the perturbations could be decoupled into two independent modes, namely polar (or even) and axial (or odd). These modes are analogous to scalar and vector modes in the FLRW model. Besides, no further decomposition into tensor modes is possible as non-trivial symmetric, transverse and trace-free rank-2 tensors cannot exist on $ S^2 $. Their equations, restricted to the FLRW metrics deformed by axial modes, agreed with those of \cite{M1, M2}. The paper \cite{CL2} numerically solves the master equation of polar waves in LTB dust model. The gauge problem plaguing the linear perturbations in cosmological models has been addressed in several papers \cite{BARDEEN, STEWART, MUKHANOV, MIEDEMA96}. The perturbations introduced in \cite{STEWART} are used to study the GWs in Kantowski-Sachs \cite{BR1} and locally rotationally symmetric class-II \cite{BR2} cosmologies.

The Bianchi spacetimes \cite{Bianchi} represent spatially homogeneous but anisotropic cosmological models \cite{EM}. The Bianchi I spacetime, having zero spatial curvature, is the simplest one, which represent the anisotropic generaliztion of a spatially flat FLRW model. The propagation of a single GW (in particular, propagation along the symmetry axis of the 2/3, 2/3, -1/3 axially-symmetric Kasner spacetime) in a Bianchi I universe has been discussed by Hu \cite{HU}. He found that the two linear polarizations decouple from one another. In \cite{MIEDEMA93} GWs have been studied within the context of a general perturbative analysis of the Bianchi I universe using the harmonic gauge, and are found to be caused by `non-material' perturbations, and therefore do not locally curve the three-dimensional hyperspaces of constant time. The usual gauge conditions have been modified in the paper \cite{CHO} and have led to the conclusion that GWs in Bianchi I universe are always transverse. Unlike the FLRW universe, where the two polarizations of the GWs decouple from one another, each separately being equivalent to a minimally coupled massless scalar field \cite{CHOref}, the two polarizations in an expanding anisotropic universe, as shown in \cite{CHO}, are coupled to one another and the coupling term depends explicitly on the choice of gauge. Moreover the GWs acquired an effective mass term because of its spin-2 field, being much more sensitive to the anisotropy than a scalar field.

Adams et al. \cite{ ADAMS1} developed a non-perturbative formalism to study GWs of arbitrary polarization, propagating through Bianchi backgrounds (type I to VII). Choosing a direction of propagation of the waves, say $z$-direction, and inserting $ z $-dependence in the line element, the symmetry in the chosen direction was broken but preserved in the transverse direction. The wave equation for the $ \psi $ `field' representing the `$+$' polarization was inhomogeneous, with the field propagating through spacetime like a massless scalar field, while the wave equation for the $ \delta $ `field' representing the `$\times$' polarization was homogeneous. They showed that only in Bianchi type I and V cosmologies, propagation of GWs in a state of either pure `+' or pure `$\times$' polarization is possible. In the other cosmological types, the spacetime structure gave a `twist' to the wave, as a result of which an initially pure `$\times$' polarization state gave rise to a `$+$' polarization component. In type III and VI cosmologies, although the GWs could propagate in a state of fixed `+' polarization, but the same did not happen for the `$\times$' polarization. Moreover, in Bianchi type II, IV, and VII cosmologies, the GWs were characterized by the presence of both polarization states. A subsequent paper \cite{ADAMS2} presented solutions to the EFEs in vacuum for a metric which is homogeneous over a spacelike hypersurface but is inhomogeneous in the perpendicular direction. They found solutions in which the inhomogeneity initially dominated the structure of the cosmic singularity and then evolved into GWs propagating over the homogeneous background.

Pereira et al. \cite{PER} have studied perturbations on Bianchi I spacetime with a minimally coupled scalar field representing an inflationary phase. 
After constructing gauge-invariant variables and choosing Newtonian gauge, they obtained the evolution equations for the gauge-invariant variables. The equations of motion involved the directional wave-number due to the anisotropy of the background spacetime, and the scalar and tensor perturbations were coupled through a non-diagonal mass term. 

The paper \cite{PELO} considered perturbations in a Bianchi I model with residual isotropy in two spatial directions, and with one scalar inflaton field. Of the three physical modes, two are coupled to each other at the linearized level. When the background becomes isotropic at late times, one of these two modes becomes the scalar perturbation, while the other one becomes the tensor perturbation, $ h_{+} $.

In this paper, we study the propagation of gravitational waves in the anisotropic but spatially homogeneous Bianchi I universe, considering the metric in spherical polar coordinates to make use of the Regge-Wheeler gauge. Previous studies on perturbations in Bianchi universes were done in Cartesian coordinates. Similar studies incorporating the Regge-Wheeler gauge have been undertaken with FLRW metric as the background \cite{M1, M2, M3, SH1, SH2}, but have not yet been applied to the Bianchi type I background. With this gauge choice, we have obtained the linearized Einstein equations for the axially perturbed Bianchi I metric in spherical polar coordinates. The metric perturbations are found to obey a wave equation. In the solution, the $ t $ and $ r $-dependences of the metric perturbations separate out as a product. The paper is organised as follows: Sec. II contains the background metric and the corresponding Einstein equations. The Regge-Wheeler gauge for axial modes is introduced in Sec. III and the perturbed field equations are derived. We solve these linearized Einstein equations to determine the two axial modes in Sec. IV, and proceed to study the field equations in presence of perfect fluid matter in Section V. The results are analyzed in Sec. VI and the summary and conclusions are presented in Sec. VII. We use an overdot to represent a derivative w.r.t $ t $ and a prime to denote a derivative w.r.t. $ r $, and have assumed geometrized units, i.e., $c=G=1$.

\section{The unperturbed Bianchi I background}
The most general metric of Bianchi I spacetime may be written as \cite{Jacobs}
\begin{equation}\label{Jacobs}
ds^2= dt^2 - \gamma_{ij}(t) dx^i dx^j,
\end{equation}
where $\gamma_{ij}$ is a symmetric $3\times3$ matrix. In spherical polar coordinates, the line element assumes the form \cite{SHAMIR1006}
\begin{equation} \label{1}
ds^2= dt^2- a^2(t) dr^2 -b^2(t) d\theta ^2 -b^2(t) \theta ^2 d\phi ^2  .
\end{equation}
The scale factors, $ a(t) $ and $ b(t) $ describe the expansion parallel and perpendicular to the radial direction respectively. In absence of matter, the Einstein equations for the metric \eqref{1} are obtained as :
\begin{equation}\label{1a}
\frac{\dot{b}}{b} \left(\frac{\dot{b}}{b} + \frac{2 \dot{a}}{a}\right) =0,
\end{equation}
\begin{equation}\label{1b}
-a^2 \left( \frac{2\ddot{b}}{b} +\frac{{\dot{b}}^2}{b^2} \right) =0,
\end{equation}
and
\begin{equation}\label{1c}
-b^2\left( \frac{\ddot{a}}{a} +\frac{\ddot{b}}{b} + \frac{\dot{a} \dot{b}}{ab}\right)  =0.
\end{equation}


Assuming that initially the scale factor $ b(t) $ is zero, i.e. $ b \left|_{(t=0)}\right. =0 $, we obtain the solution for $b(t)$ from \eqref{1b} as
\begin{equation}\label{2b}
b(t)= K t^{2/3},
\end{equation}
where $ K $ is the integration constant. For this metric, the volume expansion $\Theta$ and the shear scalar $\sigma$ are given by
\begin{equation}
 \Theta = \frac{\dot{a}}{a}+\frac{2\dot{b}}{b}, \qquad \textrm{and} \qquad  \sigma ^2 = \frac{1}{3} \left(\frac{\dot{a}}{a} -\frac{\dot{b}}{b}\right) ^2.
\end{equation}
It is known that the ratio of the shear $\sigma$ to the expansion $\Theta$ gives us a good measure of the anisotropy of a evolving universe \cite{Gron}. Cosmological models having a constant value of this ratio remain anisotropic throughout the entire evolution. Such models have been studied by a number of authors (see for example \cite{ans1, ans2, ans3, ans4}). To ensure that the anisotropy of spacetime may be sustained throughout, we incorporate this feature into the metric. In that case it may be assumed that the expansion scalar is proportional to the shear scalar, so that we have a relation of the type \cite{SHAMIR1006}
\begin{equation}\label{2c}
a=b^n,
\end{equation}
where $ n $ is an arbitrary real number and $ n \neq 0, 1 $ for non-trivial solutions. Using this relation for the metric considered here, we find that
\begin{equation}
\Theta =(n+2) \frac{\dot{b}}{b}, \qquad \textrm{and} \qquad \sigma ^2 =\frac{1}{3} \left[(n-1) \frac{\dot{b}}{b}\right] ^2.
\end{equation}
Thus
\begin{equation}
\frac{\sigma ^2}{\Theta ^2} = \frac{1}{3}\frac{(n-1)^2}{(n+2)^2}.
\end{equation}
This ratio is constant for a given $n$. Therefore, for this choice, the expansion scalar is proportional to the shear scalar in this spacetime. Exact spatially homogeneous cosmologies in which this ratio is constant, were studied in \cite{CGW}.

Inserting equation \eqref{2c} in equation \eqref{1a} and solving, we get
\begin{equation}\label{2e}
a=b^{-1/2}.
\end{equation}
This relation will be utilized in the subsequent analysis to find the solutions.

\section{The perturbed metric and field equations}
The perturbed metric can be written as
\begin{equation}\label{pert_met}
g_{\mu\nu} = g_{\mu\nu}^{(0)} + e h_{\mu\nu} + \mathcal{O}(e^2),
\end{equation}
where $ g_{\mu\nu}^{(0)} $ is the background metric \eqref{1}, $h_{\mu\nu}$ are the metric perturbations due to the GWs, $ e $ is a small parameter which gives a measure of the strength of perturbations, and $\mathcal{O}(e^2)$ represent terms involving $e^2$ or higher powers of $e$. In what follows, all terms involving higher powers of $e$ will be neglected.

In this paper, we follow the Regge-Wheeler scheme of perturbation \cite{RW} to study the propagation of GWs in Bianchi Type I universe. By generalizing the perturbations into tensor harmonics, Regge and Wheeler found explicit solutions for the differential equations satisfied by the small perturbations. The solution appeared as a product of four factors, each depending on a single coordinate $t,r,\theta,\phi$.
The symmetry of the metric allowed the angular momentum to be defined for all four coordinates. Under a rotation of the frame about the origin (rotations on $ t $=constant, $r $=constant hypersurface), the ten independent components of $ h_{\mu\nu} $ transformed in different manners. The components $ h_{00} $, $ h_{01} $, $ h_{11} $ transformed like scalars, 
($ h_{02} , h_{03} $) and ($ h_{12} , h_{13} $) transformed like vectors, 
and $ h_{22} $, $ h_{23} $ \& $ h_{33} $ transformed like a second-order tensor, when considered as covariant quantities on a sphere. These scalars, vectors and tensors were expressed in terms of spherical harmonics $ Y_{lm} $ where $ l $ is the angular momentum whose projection on the $z$ axis is $ m $. The spherical harmonics, were characterized by distinct parities, either ``odd'' or ``even''. The grouping of the terms of odd (or even) parity yielded the ``odd'' (or ``even'') perturbation matrices. The angular dependence of the perturbations were explicitly specified by these matrices. Since all values of $m$ led to the same radial equation, therefore the value $m=0$ was chosen, so that the $ \phi $-dependence disappeared. In spite of that, the axial waves were composed of three unknown functions of $r$, and the polar waves contained seven unknown functions. The introduction of the `Regge-Wheeler' gauge yielded the canonical form of the axial and polar waves.

In this paper, we will restrict our analysis to the odd or axial wave perturbations. In the Regge-Wheeler gauge \cite{RW}, there are only two non-zero components of $ h_{\mu\nu} $ in the case of axial waves. These are
\begin{equation}\label{nonzero_pert}
h_{t \phi} =h_0(t,r) \sin \theta (\partial_{\theta}Y) \hspace{0.2cm} \text {and} \hspace{0.2cm}
h_{r \phi} =h_1(t,r) \sin \theta (\partial_{\theta}Y).
\end{equation}
Here, the spherical harmonics $ Y_{lm}(\theta, \phi) $ are denoted by $Y$, and $m=0$ is chosen.  Also $ l \geq 2 $ for wavelike solutions \cite{SH1}. Besides, we know that $ Y_{lm}(\theta, \phi) $ satisfies the relation
\begin{equation}\label{sph_harc}
\partial_{\theta} \partial_{\theta} Y = -l(l+1)Y - \cot\theta (\partial_{\theta} Y).
\end{equation}

The axial perturbations for the background metric \eqref{1} are given by
\begin{equation}\label{3}
ds^2= dt^2- a^2(t) dr^2 -b^2(t) d\theta ^2 -b^2(t) \theta ^2 d\phi ^2 
+2eh_0(t,r) \sin \theta (\partial_{\theta}Y) dtd\phi +2eh_1(t,r) \sin \theta (\partial_{\theta}Y) drd\phi + \mathcal{O}(e^2).
\end{equation}

The equations \eqref{1a}, \eqref{1b} and \eqref{1c} 
hold for the perturbed metric \eqref{3} as well. Retaining only the first order terms in $e$, the additional linearized Einstein equations arising due to the perturbations in vacuum are as follows:

\begin{equation}\label{3a}
\frac{e}{2} \sin\theta (\partial_{\theta}Y) \left[
\frac{h_0''}{a^2} -\frac{\dot{h}_1'}{a^2}  +\frac{2 \dot{b}h_1'}{a^2 b} +\frac{2 \ddot{a}h_0}{a} +\frac{2 \ddot{b}h_0}{b}
+\frac{2 \dot{a} \dot{b} h_0}{ab}  -\frac{h_0}{b^2 } \left\lbrace  l(l+1) \right\rbrace  \right] =0,
\end{equation}

\begin{equation}\label{3b}
-\frac{e}{2} \sin\theta (\partial_{\theta}Y) \left[
\ddot{h}_1 -\dot{h}_0' -\frac{\dot{a}\dot{h}_1}{a}  +\frac{\dot{a}h_0'}{a} -\frac{2 \dot{b} h_0'}{b} -\frac{2\ddot{a} h_1}{a}
-\frac{4\ddot{b} h_1}{b} -\frac{2 \dot{b}^2 h_1}{b^2} +\frac{h_1}{b^2} \left\lbrace  l(l+1)
\right\rbrace   \right] =0,
\end{equation}
and
\begin{equation}\label{3c}
 \frac{e}{2} \left( \dot{h}_0+ \frac{\dot{a}h_0}{a} -\frac{h_1'}{a^2} \right) \left( \cos\theta (\partial_\theta Y)-\frac{2\sin \theta(\partial_\theta Y)}{\theta} +\sin\theta (\partial_\theta \partial_\theta Y) \right) =0.
\end{equation}

For the sake of simplicity, we will assume that $\theta$ is small, so that $\tan \theta = \theta$ and $ \cot \theta =1/\theta $, in deriving the last term of both the equations \eqref{3a} and \eqref{3b}. Consistency of \eqref{3c} with \eqref{sph_harc} demands that we should have
\begin{equation}\label{3d}
\dot{h}_0 +\frac{\dot{a}h_0}{a} -\frac{h_1'}{a^2} =0
\hspace{0.3cm} \Rightarrow {h_1'}= a^2\dot{h}_0+ a\dot{a}h_0.
\end{equation}

\section{Axial gravitational waves in Bianchi I universe}

We now proceed to solve the above equations to find the expressions for the perturbations $ h_0(t,r) $ and $ h_1(t,r) $.
Using equations \eqref{1c}, \eqref{2e}, and \eqref{3d}, 
the equation \eqref{3a} can be expressed completely in terms of $ h_0(t,r) $ and we obtain

\begin{equation}\label{4c}
 - \ddot{h}_0 +b h_0''  +\frac{7\dot{b}}{2b} \dot{h}_0 -\frac{2{\dot{b}}^2}{b^2} h_0  + \frac{\ddot{b}}{2b}  h_0
-\frac{1}{b^2} l(l+1)h_0 =0. 
\end{equation}

Let us now define a new quantity $ \mathcal{Q}(t,r) $ such that
\begin{equation}\label{G1}
h_0(t,r)= r^\alpha (b(t))^\beta \mathcal{Q}(t,r).
\end{equation}

Inserting the expression of $ b(t) $ from equation \eqref{2b}, and also choosing  $ l=2 $, equation \eqref{4c} reduces to the form
\begin{equation} \label{G2}
\begin{split}
- K^\beta t^{\frac{2\beta}{3}} r^\alpha \ddot{\mathcal{Q}} + K^{\beta +1} t^{\frac{2}{3}(\beta +1)} r^\alpha \mathcal{Q}''  + \left[ \left( -\frac{4}{3}\beta + \frac{7}{3}\right) K^\beta t^{(\frac{2\beta}{3}-1)} r^\alpha \right]  \dot{\mathcal{Q}}
+ 2\alpha K^{\beta +1} t^{\frac{2}{3}(\beta +1)} r^{\alpha -1}  \mathcal{Q}' \\
+ \left[ \alpha^2 K^{\beta +1}  t^{\frac{2}{3}(\beta +1)} r^{\alpha -2}  -  \alpha K^{\beta +1} t^{\frac{2}{3}(\beta +1)} r^{\alpha -2}  +  \left( \frac{20}{9} \beta -  \frac{4}{9} \beta^2 -1 \right) K^\beta t^{(\frac{2\beta}{3}-2)} r^\alpha
\right. \\  \left.
- 6 K^{\beta-2} t^{(\frac{2\beta}{3}-\frac{4}{3})} r^\alpha \right] \mathcal{Q}  =0,
\end{split}
\end{equation}
where $K$ is a constant. Equation \eqref{G2} is a wave equation in $ \mathcal{Q}(t,r) $. It describes one of the two gravitational modes and can be solved independently of the other equations. The solution is obtained by employing the method of separation of variables by assuming that
\begin{equation}\label{6a}
\mathcal{Q}(t,r)= \mathcal{T}(t) \mathcal{R}(r),
\end{equation}
so that \eqref{G2} leads to two differential equations with the separation constant $ -p^2 $ :
\begin{equation}\label{G3}
\ddot{\mathcal{T}} + \left( \frac{4\beta}{3} - \frac{7}{3}\right)  \frac{\dot{\mathcal{T}}}{t} +\left( -\frac{20\beta}{9t^{2}}  + \frac{4\beta^2}{9t^{2}} + \frac{1}{t^{2}} +\frac{6}{K^2 t^{4/3}} + p^2 K t^{2/3} \right)\mathcal{T}  =0,
\end{equation}
and
\begin{equation}\label{G4}
{\mathcal{R}''} +  \frac{2\alpha \mathcal{R}'}{r} +\left( \frac{\alpha^2}{r^2} - \frac{\alpha}{r^2} + p^2 \right) \mathcal{R} =0.
\end{equation}

The solutions are respectively \cite{Maple}:
\begin{equation}\label{G5}
\mathcal{T}(t) = c_1 t^{(3- 2\beta/3)} {\exp(\xi t^{4/3})} \Pi  +c_2 t^{(3- 2\beta/3)} \exp(\xi t^{4/3})
 \Pi  \left( \int \dfrac{\exp(-2\xi t^{4/3}) dt }{t^{11/3} \Pi^2} \right) ,
\end{equation}
and
\begin{equation}\label{G6}
\mathcal{R}(r) = r^{-\alpha} \left( c_3 \sin pr + c_4 \cos pr \right).
\end{equation}
Here, $ c_1 $, $ c_2 $, $ c_3 $ and $ c_4 $ are the integration constants, and $ \xi =\dfrac{3}{4} \sqrt{-K p^2} $.

The quantity $ \Pi =\text{HeunB}\left( 4, 0, 0, \dfrac{9\sqrt{6}i}{K^2 (-Kp^2)^{1/4}}, \dfrac{\sqrt{6}i}{2} (-K p^2)^{1/4} t^{2/3} \right) $ represents the Heun function which is the solution of Heun's biconfluent equation \cite{HOR, RON}.

Therefore, $ \mathcal{Q}(t,r) $ is now completely determined in terms of $ \mathcal{T}(t) $ and $ \mathcal{R}(r) $ which are evaluated in \eqref{G5} and \eqref{G6}. Plugging these quantities into equation \eqref{G1} and using \eqref{2b}, we obtain the expression for $ h_0(t,r) $ as follows: 
\begin{equation}\label{G7a}
h_0(t,r)= r^\alpha K^\beta  t^{2\beta/3} \mathcal{Q}(t,r).
\end{equation}

Finally, $ h_1(t,r) $ is determined from \eqref{3d} and is given by
\begin{equation}\label{G8}
h_1(t,r)=  f(t)+ \frac{1}{K t^{2/3}} \int_{r_0}^{r}\left(   \dot{h}_0(t,r) - \frac{h_0(t,r)}{3t} \right) dr ,
\end{equation}
where $ f(t) $ is the integration constant, and $ r_0 $ characterizes the initial hypersurface from which the gravitational waves emanate. The function $ f(t) $ is arbitrary, and can be chosen to be zero, i.e. $ f(t) = 0$, as a result of which we have
\begin{equation}\label{G9}
h_1(t,r) = K^{\beta-1}\left(  t^{(\frac{2\beta}{3}-\frac{2}{3})} \dot{\mathcal{T}} + \left( \frac{2}{3} \beta -\frac{1}{3}\right) t^{(\frac{2\beta}{3}-\frac{5}{3})} \mathcal{T} \right) \int_{r_0}^{r}  r^{\alpha} \mathcal{R} dr .
\end{equation}

To develop our understanding of the effect of the advancing GWs in this spacetime, we now consider a few cases for particular values of $\alpha$ and $\beta$ in \eqref{G1}, which are presented hereunder.

\subsection{Case 1}

Let us set  $\alpha=1$ and $\beta=2$ in \eqref{G1} so that 
\begin{equation} \label{5a1}
h_0(t,r)= r b(t)^2 \mathcal{Q}(t,r).
\end{equation}
Thus equation \eqref{4c} reads
\begin{equation} \label{5b1}
-r b^2 \ddot{\mathcal{Q}} +r b^3 \mathcal{Q}'' -\frac{1}{2}r b\dot{b} \dot{\mathcal{Q}} +2b^3 \mathcal{Q}'  -\frac{3}{2}r b \ddot{b} \mathcal{Q} +3r \dot{b}^2 \mathcal{Q} -r l(l+1) \mathcal{Q} =0.
\end{equation}
Using \eqref{2b}, and choosing  $ l=2 $, equation \eqref{5b1} can be reduced to the form
\begin{equation} \label{5c1}
- K^2 r t^{4/3} \ddot{\mathcal{Q}} + K^3 r t^{2} \mathcal{Q}'' -\frac{1}{3} K^2 r t^{1/3} \dot{\mathcal{Q}} +2K^3 t^{2} \mathcal{Q}' +\frac{5}{3} K^2 r t^{-2/3} \mathcal{Q} -6 r  \mathcal{Q} =0,  
\end{equation}
where the constant $K$ comes from \eqref{2b}. The wave equation \eqref{5c1} is solved once again by the method of separation of variables, and we obtain  two differential equations with the separation constant $ -m_1^2 $ :
\begin{equation}\label{6b1}
\ddot{\mathcal{T}} +\frac{1}{3} \frac{\dot{\mathcal{T}}}{t} +\left( -\frac{5}{3t^{2}} +\frac{6}{K^2 t^{4/3}} + m_1^2 K t^{2/3} \right)\mathcal{T}  =0,
\end{equation}
\begin{equation}\label{6c1}
\text{and} \hspace{0.5cm}  {\mathcal{R}''} +  \frac{2\mathcal{R}'}{r} +m_1^2 {\mathcal{R}} =0.
\end{equation}

The solutions are respectively:
\begin{equation}\label{6d1}
\mathcal{T}(t) = k_1 t^{5/3} {\exp(\xi t^{4/3})} \Pi  +k_2 t^{5/3} \exp(\xi t^{4/3})
 \Pi  \left( \int \dfrac{\exp(-2\xi t^{4/3}) dt }{t^{11/3} \Pi^2} \right) ,
\end{equation}
\begin{equation}\label{6e1}
\text{and} \hspace{0.5cm}  \mathcal{R}(r) = \frac{1}{r} \left( k_3 \cos m_1 r + k_4 \sin m_1 r \right) .
\end{equation}
Here, $ k_1 $, $ k_2 $, $ k_3 $ and $ k_4 $ are the integration constants, and the function $\xi$ is given by $ \xi =\dfrac{3}{4} \sqrt{-K m_1^2} $. The Heun function is given by $ \Pi =\text{HeunB}\left( 4, 0, 0, \dfrac{9\sqrt{6}i}{K^2 (-Km_1^2)^{1/4}}, \dfrac{\sqrt{6}i}{2} (-K m_1^2)^{1/4} t^{2/3} \right)  $.

The expression for $ \mathcal{Q}(t,r) $ is obtained in terms of $ \mathcal{T}(t) $ and $ \mathcal{R}(r) $, and $ h_0(t,r) $ is now given by 
\begin{equation}\label{7a}
h_0(t,r)= K^2 r t^{4/3} \mathcal{Q}(t,r).
\end{equation}
From \eqref{3d}, we obtain $ h_1(t,r) $ as:
\begin{equation}\label{7b1}
h_1(t,r)=   f_1(t)+ \frac{1}{K t^{2/3}} \int_{r_0}^{r}\left(   \dot{h}_0(t,r) - \frac{h_0(t,r)}{3t} \right) dr ,
\end{equation}
where $ f_1(t) $ is the integration constant, and $ r_0 $ characterizes the initial hypersurface giving rise to gravitational waves. Setting the arbitrary constant $ f_1(t) $ to zero, we obtain
\begin{equation}\label{7c1}
h_1(t,r) = K ( t^{2/3} \dot{\mathcal{T}} +t^{-1/3} \mathcal{T} ) \int_{r_0}^{r}  r \mathcal{R} dr
= K ( t^{2/3} \dot{\mathcal{T}} +t^{-1/3} \mathcal{T} ) \left[  \frac{1}{m_1} \left(k_3 \sin m_1r - k_4 \cos m_1r \right) \right]_{r_0}^{r} .
\end{equation}

\subsection{Case 2}

For $\alpha=0$ and $\beta=2$ in \eqref{G1}, we have  
\begin{equation} \label{8a}
h_0(t,r)= b(t)^2 \mathcal{Q}(t,r).
\end{equation}
Simplifying as before and choosing  $ l=2 $, we are led to the differential equation
\begin{equation}\label{8c}
-K^2 t^{4/3} \ddot{\mathcal{Q}} +K^3 t^{2} \mathcal{Q}'' -\frac{1}{3} K^2 t^{1/3} \dot{\mathcal{Q}} +\frac{5}{3} K^2 t^{-2/3} \mathcal{Q} -6 \mathcal{Q} =0,
\end{equation}
where $K$ is given by \eqref{2b}. Using the method of separation of variables as per assumption \eqref{6a}, equation \eqref{8c} leads to two differential equations with the separation constant $ -m_2^2 $ :
\begin{equation}\label{9a}
\ddot{\mathcal{T}} +\frac{1}{3} \frac{\dot{\mathcal{T}}}{t} +\left( -\frac{5}{3t^{2}} +\frac{6}{K^2 t^{4/3}} + m_2^2 K t^{2/3} \right)\mathcal{T}  =0,
\end{equation}
\begin{equation}\label{9b}
\text{and} \hspace{0.5cm}  {\mathcal{R}''}  +m_2^2 {\mathcal{R}} =0.
\end{equation}

The solutions are respectively:
\begin{equation}\label{9c}
\mathcal{T}(t) = p_1 t^{5/3} {\exp(\xi t^{4/3})} \Pi  + p_2 t^{5/3} \exp(\xi t^{4/3})
 \Pi  \left( \int \dfrac{\exp(-2\xi t^{4/3}) dt }{t^{11/3} \Pi^2} \right) ,
\end{equation}
\begin{equation}\label{9d}
\text{and} \hspace{0.5cm}  \mathcal{R}(r) = p_3 \cos m_2 r + p_4 \sin m_2 r .
\end{equation}
Here $ p_1 $, $ p_2 $, $ p_3 $ and $ p_4 $ are the integration constants, and $ \xi =\dfrac{3}{4} \sqrt{-K m_2^2} $.
The Heun function is given by $ \Pi =\text{HeunB}\left( 4, 0, 0, \dfrac{9\sqrt{6}i}{K^2 (-Km_2^2)^{1/4}}, \dfrac{\sqrt{6}i}{2} (-K m_2^2)^{1/4} t^{2/3} \right)  $.

Finally, the expression for $ h_0(t,r) $ is 
\begin{equation}\label{10a}
h_0(t,r)= K^2 t^{4/3} \mathcal{Q}(t,r),
\end{equation}
and $ h_1(t,r) $ is obtained from equation \eqref{3d} as
\begin{equation}\label{10b}
h_1(t,r)=   f_2(t)+ \frac{1}{K t^{2/3}} \int_{r_0}^{r}\left(   \dot{h}_0(t,r) - \frac{h_0(t,r)}{3t} \right) dr ,
\end{equation}
where $ f_2(t) $ is the integration constant, and $ r_0 $ characterizes the initial hypersurface giving rise to gravitational waves. Once again the arbitrary function $ f_2(t) $ is set to zero (i.e. $ f_2(t) = 0$), and we have
\begin{equation}\label{10c}
h_1(t,r) = K ( t^{2/3} \dot{\mathcal{T}} +t^{-1/3} \mathcal{T} ) \int_{r_0}^{r}   \mathcal{R} dr
= K ( t^{2/3} \dot{\mathcal{T}} +t^{-1/3} \mathcal{T} ) \left[\frac{1}{m_2} (p_3 \sin m_2 r - p_4 \cos m_2r) \right]_{r_0}^{r}.
\end{equation}

\subsection{Case 3}

For the case $\alpha=0$ and $\beta=1$, we have 
\begin{equation} \label{11a}
h_0(t,r)= b(t) \mathcal{Q}(t,r).
\end{equation}
Thus equation \eqref{G2} now reads
\begin{equation}\label{11c}
-K t^{2/3} \ddot{\mathcal{Q}} + K^2 t^{4/3} \mathcal{Q}'' + K t^{-1/3} \dot{\mathcal{Q}} + \frac{7}{9} K t^{-4/3} \mathcal{Q} -\frac{6}{K} t^{-2/3} \mathcal{Q} =0,
\end{equation}
where we have used \eqref{2b}, and chosen $ l=2 $, with $K$ as the constant. Assuming $\mathcal{Q}$ to be given by \eqref{6a}, we arrive at two differential equations with the separation constant $ -m_3^2 $:
\begin{equation}\label{12a}
\ddot{\mathcal{T}} - \frac{\dot{\mathcal{T}}}{t} +\left( -\frac{7}{9t^{2}} +\frac{6}{K^2 t^{4/3}} +	 m_3^2 K t^{2/3} \right)\mathcal{T}  =0,
\end{equation}
\begin{equation}\label{12b}
\text{and} \hspace{0.5cm}  {\mathcal{R}''}  +m_3^2 {\mathcal{R}} =0.
\end{equation}

The solutions are respectively:
\begin{equation}\label{12c}
\mathcal{T}(t) = q_1 t^{7/3} {\exp(\xi t^{4/3})} \Pi  + q_2 t^{7/3} \exp(\xi t^{4/3})
 \Pi  \left( \int \dfrac{\exp(-2\xi t^{4/3}) dt }{t^{11/3} \Pi^2} \right) ,
\end{equation}
\begin{equation}\label{12d}
\text{and} \hspace{0.5cm}  \mathcal{R}(r) = q_3 \cos m_3 r + q_4 \sin m_3 r ,
\end{equation}
where $ q_1 $, $ q_2 $, $ q_3 $ and $ q_4 $ are the integration constants, $ \xi =\dfrac{3}{4} \sqrt{-K m_3^2} $, and the Heun function is $ \Pi =\text{HeunB}\left( 4, 0, 0, \dfrac{9\sqrt{6}i}{K^2 (-Km_3^2)^{1/4}}, \dfrac{\sqrt{6}i}{2} (-K m_3^2)^{1/4} t^{2/3} \right)  $.

The expression for $ h_0(t,r) $ now stands as 
\begin{equation}\label{13a}
h_0(t,r)= K t^{2/3} \mathcal{Q}(t,r).
\end{equation}

From equation \eqref{3d} we find that
\begin{equation}\label{13b}
h_1(t,r)=   f_3(t)+ \frac{1}{K t^{2/3}} \int_{r_0}^{r}\left(   \dot{h}_0(t,r) - \frac{h_0(t,r)}{3t} \right) dr ,
\end{equation}
where $ f_3(t) $ is the integration constant. Setting $ f_3(t) = 0$ we obtain
\begin{equation}\label{13c}
h_1(t,r) = \left( \dot{\mathcal{T}} +\frac{1}{3t} \mathcal{T} \right) \int_{r_0}^{r}  \mathcal{R} dr
=  \left( \dot{\mathcal{T}} +\frac{1}{3t}  \mathcal{T} \right)  \left[\frac{1}{m_3} (q_3 \sin m_3 r - q_4 \cos m_3 r) \right]_{r_0}^{r}.
\end{equation}

\subsection{Case 4}

The last example that we consider, corresponds to the case $\alpha=0$ and $\beta=3$, so that 
\begin{equation} \label{14a}
h_0(t,r)= b(t)^3 \mathcal{Q}(t,r).
\end{equation}
Simplifying as before and choosing  $ l=2 $, we are led to the following wave equation for $\mathcal{Q}$:
\begin{equation}\label{14c}
-K^3 t^{2} \ddot{\mathcal{Q}} +K^4 t^{8/3} \mathcal{Q}'' -\frac{5}{3} K^3 t \dot{\mathcal{Q}} +\frac{5}{3} K^3  \mathcal{Q} -6K t^{2/3} \mathcal{Q} =0.
\end{equation}
Employing the method of separation of variables with the assumption \eqref{6a}, the equation \eqref{14c} separates into two differential equations with the separation constant $ -m_4^2 $:
\begin{equation}\label{15a}
\ddot{\mathcal{T}} +\frac{5}{3} \frac{\dot{\mathcal{T}}}{t} +\left( -\frac{5}{3t^{2}} +\frac{6}{K^2 t^{4/3}} + m_4^2 K t^{2/3} \right)\mathcal{T}  =0,
\end{equation}
\begin{equation}\label{15b}
\text{and} \hspace{0.5cm}  {\mathcal{R}''}  + m_4^2 {\mathcal{R}} =0.
\end{equation}

The solutions are respectively:
\begin{equation}\label{15c}
\mathcal{T}(t) = w_1 t \exp(\xi t^{4/3}) \Pi  + w_2 t \exp(\xi t^{4/3})
 \Pi  \left( \int \dfrac{\exp(-2\xi t^{4/3}) dt }{t^{11/3} \Pi^2} \right) ,
\end{equation}
\begin{equation}\label{15d}
\text{and} \hspace{0.5cm}  \mathcal{R}(r) = w_3 \cos m_4 r + w_4 \sin m_4 r .
\end{equation}
Here, $ w_1 $, $ w_2 $, $ w_3 $ and $ w_4 $ are the integration constants, and $ \xi =\dfrac{3}{4} \sqrt{-K m_4^2} $.
The Heun function is given by $ \Pi =\text{HeunB}\left( 4, 0, 0, \dfrac{9\sqrt{6}i}{K^2 (-Km_4^2)^{1/4}}, \dfrac{\sqrt{6}i}{2} (-K m_4^2)^{1/4} t^{2/3} \right)  $.

Using equation \eqref{14a} and equation \eqref{2b}, we get
\begin{equation}\label{16a}
h_0(t,r)= K^3 t^{2} \mathcal{Q}(t,r),
\end{equation}
and from equation \eqref{3d}, we have
\begin{equation}\label{16b}
h_1(t,r)=   f_4(t)+ \frac{1}{K t^{2/3}} \int_{r_0}^{r}\left(   \dot{h}_0(t,r) - \frac{h_0(t,r)}{3t} \right) dr ,
\end{equation}
where $ f_4(t) $ is the arbitrary constant of integration, which is set to zero, so that we obtain the following expression for $h_1(t,r)$:
\begin{equation}\label{16c}
h_1(t,r) = K^2 \left(t^{4/3} \dot{\mathcal{T}} +\frac{5}{3} t^{1/3} \mathcal{T} \right) \int_{r_0}^{r}  \mathcal{R} dr
= K^2 \left(t^{4/3} \dot{\mathcal{T}} +\frac{5}{3} t^{1/3} \mathcal{T} \right)  \left[\frac{1}{m_4} (w_3 \sin m_4 r - w_4 \cos m_4 r) \right]_{r_0}^{r},
\end{equation}
where $r_0$ characterizes the initial hypersurface giving rise to gravitational waves.

\subsection{Comments}
In the above calculations, we have assumed that $ \cot \theta =1/ \theta $, which is true for small values of $\theta$. If $ \cot \theta \neq 1/ \theta $, then there appears an additional term with $ \theta $-dependence in equations \eqref{3a} and \eqref{3b}. These equations are then given by
\begin{equation}\label{3e}
\begin{split}
\frac{e}{2} \sin\theta (\partial_{\theta}Y) \left[
\frac{h_0''}{a^2} -\frac{\dot{h}_1'}{a^2}  +\frac{2 \dot{b}h_1'}{a^2 b} +\frac{2 \ddot{a}h_0}{a} +\frac{2 \ddot{b}h_0}{b}
+\frac{2 \dot{a} \dot{b} h_0}{ab} -\frac{h_0}{b^2 } \left\lbrace  l(l+1) \right\rbrace
\right.  \\ \left.
+\frac{h_0}{b^2} \left\lbrace  l(l+1)\frac{Y }{(\partial_{\theta}Y)} \left(-\cot \theta +\frac{1}{\theta}\right)  \right\rbrace \right]=0,
\end{split}
\end{equation}
and
\begin{equation}\label{3f}
\begin{split}
-\frac{e}{2} \sin\theta (\partial_{\theta}Y) \left[
\ddot{h}_1 -\dot{h}_0' -\frac{\dot{a}\dot{h}_1}{a}  +\frac{\dot{a}h_0'}{a} -\frac{2 \dot{b} h_0'}{b} -\frac{2\ddot{a} h_1}{a}
-\frac{4\ddot{b} h_1}{b} - \frac{2 \dot{b}^2 h_1}{b^2} +\frac{h_1}{b^2} \left\lbrace  l(l+1) \right\rbrace
\right.  \\ \left.
-\frac{h_1}{b^2} \left\lbrace  l(l+1)\frac{Y }{(\partial_{\theta}Y)} \left(-\cot \theta +\frac{1}{\theta}\right)  \right\rbrace
\right] =0.
\end{split}
\end{equation}
We want to point out that in this analysis we have solved equation \eqref{3a} (or its variant) by the method of separation of variables after choosing the expression for $ h_0(t,r) $ in terms of $ \mathcal{Q}(t,r) $. Since these quantities are independent of $ \theta $, it is not possible to extract a separate equation for $ \theta $ from equation \eqref{3e} or equation \eqref{3f}.

Further, we have assumed that $ l \geq 2 $ for wavelike solutions \cite{SH1}. The case $ l=0 $ for spherical harmonics corresponds to spherical symmetry. Higher values of $ l $ indicate departure from spherical symmetry, and lead to non-zero quadrupole moment. We also know that GWs are associated with quadrupole moment. According to Clarkson et al. \cite{CL1}, scalars on $ S^{2} $ can be expressed as a sum over polar modes, and higher-rank tensors as sums over both the polar and axial modes. Only a scalar function can contain spherical perturbations characterised by $ l = 0 $. A dipole term ($ l = 1 $) appears in the expansion of scalars and vectors in terms of spherical harmonics. Higher multipoles ($ l \geq 2 $) can be present in tensors of any rank. For the axial modes coming from the expansion of both vector and tensor functions, we therefore, need to consider $ l\geq 2 $. The height of the effective potential barrier, given by the coefficient of $ \mathcal{Q}(t,r) $ in the master differential equation, depends on the values of $ l $ \cite{RW}. In our case, different values of $ l(\geq 2) $ will lead to slightly different solutions of the $ \mathcal{T}(t) $ equations (as the Heun's functions are different), and hence $ \mathcal{Q}(t,r) $ will change, thereby affecting both $ h_0(t,r) $ and $ h_1(t,r) $, although there will be no change in the $ \mathcal{R}(r) $ solution. For a different value of $l$, for example $ l=3 $, assuming the variable separation in (\ref{G1}), with the separation constant $ -p^2 $, we obtain
\begin{equation}\label{2}
\ddot{\mathcal{T}} + \left( \frac{4\beta}{3} - \frac{7}{3}\right)  \frac{\dot{\mathcal{T}}}{t} +\left( -\frac{20\beta}{9t^{2}}  + \frac{4\beta^2}{9t^{2}} + \frac{1}{t^{2}} +\frac{12}{K^2 t^{4/3}} + p^2 K t^{2/3} \right)\mathcal{T}  =0,
\end{equation}
and
\begin{equation}\label{3aa}
{\mathcal{R}''} +  \frac{2\alpha \mathcal{R}'}{r} +\left( \frac{\alpha^2}{r^2} - \frac{\alpha}{r^2} + p^2 \right) \mathcal{R} =0,
\end{equation}
with the corresponding solutions :
\begin{equation}\label{4}
\mathcal{T}(t) = \bar{c}_1 t^{(3- 2\beta/3)} {\exp(\xi t^{4/3})} \Pi  + \bar{c}_2 t^{(3- 2\beta/3)} \exp(\xi t^{4/3})
 \Pi  \left( \int \dfrac{\exp(-2\xi t^{4/3}) dt }{t^{11/3} \Pi^2} \right) ,
\end{equation}
and
\begin{equation}\label{5}
\mathcal{R}(r) = r^{-\alpha} \left( \bar{c}_3 \sin pr + \bar{c}_4 \cos pr \right),
\end{equation}
respectively, where, $ \bar{c}_1 $, $ \bar{c}_2 $, $ \bar{c}_3 $ and $ \bar{c}_4 $ are the integration constants, \hspace{0.1cm} $ \xi =\dfrac{3}{4} \sqrt{-K p^2} $, and $ \Pi =\text{HeunB}\left( 4, 0, 0, \dfrac{18\sqrt{6}i}{K^2 (-Kp^2)^{1/4}}, \dfrac{\sqrt{6}i}{2} (-K p^2)^{1/4} t^{2/3} \right)  $ represents the Heun function. Comparing \eqref{5} with \eqref{G6}, we can say that the nature of variation of $\mathcal{R}$ remains the same, although the Heun's function in \eqref{4} is now different.

\section{Field equations in the presence of matter}

If $u^\alpha$ is the fluid four-velocity, $ \rho $ the energy density and $ p $ the pressure, then the energy-momentum tensor of a perfect fluid is given by
\begin{equation}
T_{\mu\nu}= (\rho+p) u_{\mu}u_{\nu} -p g_{\mu\nu}.
\end{equation}
For the Bianchi I universe given by \eqref{1}, with matter content as a perfect fluid, the background field equations may be obtained easily, where we use the subscript `0' to represent the parameters of the background metric:
\begin{equation}\label{1ar}
\frac{2 \dot{a}\dot{b}}{ab} + \frac{\dot{b}^2}{b^2} =\rho_0,
\end{equation}
\begin{equation}\label{1br}
-\left( \frac{2\ddot{b}}{b} +\frac{{\dot{b}}^2}{b^2} \right) =p_0,
\end{equation}
and
\begin{equation}\label{1cr}
-\left( \frac{\ddot{a}}{a} +\frac{\ddot{b}}{b} +\frac{\dot{a} \dot{b}}{ab} \right) =p_0.
\end{equation}
For the axially perturbed metric \eqref{3} in presence of matter, the perturbed energy density and pressure of the fluid can be obtained as in the FRLW case \cite{M2, SH1}:
\begin{equation}\label{3ar}
\rho =\rho_0 (1+e\cdot \Delta(t,r) Y) +\mathcal{O}(e^2),
\end{equation}
\begin{equation}\label{3br}
p =p_0 (1+e\cdot \Psi(t,r) Y) +\mathcal{O}(e^2).
\end{equation}
where $ \Delta(t,r) $ and $ \Psi(t,r) $ are the perturbations in the energy density and pressure respectively. The terms $ \Delta $ and $ \Psi $ are related because the background energy density $ \rho_0 $ and pressure $ p_0 $ are related by a suitable equation of state.  Moreover, the fluid may or may not be comoving with the unperturbed cosmological expansion of the universe. As a result, we have to simultaneously consider the perturbations in its four-velocity \cite{M2, SH1}. The perturbations in the components of the fluid four-velocity $ u_\alpha =(u_0,u_1,u_2,u_3) $ may be defined for the metric \eqref{3} in the following way, where the subscript `0' indicate parameters for the background spacetime:
\begin{equation}\label{3cr}
u_0 = \frac{2 g_{00}^{(0)} +e h_{00}}{2} +\mathcal{O}(e^2),
\end{equation}
\begin{equation}\label{3dr}
u_1 = e a(t) w(t,r) Y +\mathcal{O}(e^2),
\end{equation}
\begin{equation}\label{3er}
u_2 = e v(t,r) (\partial_{\theta}Y) +\mathcal{O}(e^2),
\end{equation}
\begin{equation}\label{3fr}
u_3 = e u(t,r) \sin \theta (\partial_{\theta}Y) +\mathcal{O}(e^2).
\end{equation}
These four-velocity components satisfy the relation: $ u_{\mu}u^{\mu} =1 +\mathcal{O}(e^2) $.
For axial waves, we have $ h_{00}=0. $ The linearised field equations for the perturbed metric  \eqref{3} in presence of matter are as follows :
\begin{equation}\label{4ar}
\frac{2 \dot{a}\dot{b}}{ab} + \frac{\dot{b}^2}{b^2} =\rho_0 (1+e\Delta Y),
\end{equation}
\begin{equation}\label{4br}
-\left( \frac{2\ddot{b}}{b} +\frac{{\dot{b}}^2}{b^2} \right) = p_0(1+e\Psi Y),
\end{equation}
\begin{equation}\label{4cr}
-\left( \frac{\ddot{a}}{a} +\frac{\ddot{b}}{b} +\frac{\dot{a} \dot{b}}{ab} \right) = p_0(1+e\Psi Y),
\end{equation}
\begin{equation}\label{4er}
(\rho_0 +p_0) e a w Y =0,
\end{equation}
\begin{equation}\label{4fr}
(\rho_0 +p_0) e v (\partial_{\theta}Y) =0,
\end{equation}
\begin{equation}\label{5er}
\begin{split}
\frac{h_0''}{a^2} -\frac{\dot{h}_1'}{a^2}  +\frac{2 \dot{b}h_1'}{a^2 b}
-\frac{h_0}{b^2} \left\lbrace  l(l+1) \right\rbrace  = 2u(\rho_0 +p_0),
\end{split}
\end{equation}
\begin{equation}\label{5fr}
\begin{split}
\ddot{h}_1 -\dot{h}_0' -\frac{\dot{a}\dot{h}_1}{a}  +\frac{\dot{a}h_0'}{a} -\frac{2 \dot{b} h_0'}{b} -\frac{2\ddot{a} h_1}{a}
 +\frac{h_1}{b^2} \left\lbrace  l(l+1) \right\rbrace  = 0,
\end{split}
\end{equation}
\begin{equation}\label{5gr}
\dot{h}_0+ \frac{\dot{a}h_0}{a} -\frac{h_1'}{a^2} =0.
\end{equation}
The assumption in the vacuum case that $ \theta $ is small so that $ \tan \theta =\theta $, and $ \cot \theta =1/ \theta $, is used here as well (to derive equations \eqref{5er},\eqref{5fr} and \eqref{5gr}, where we have also utilized equations \eqref{1ar}-\eqref{1cr}). The perturbation equations \eqref{4ar}-\eqref{4cr} can also be simplified using the background field equations. From equations \eqref{1ar} and \eqref{4ar}, we get
\begin{equation}\label{5ar}
\Delta \cdot \rho_0 =0.
\end{equation}
Combining \eqref{1br} and \eqref{1cr} with equations \eqref{4br} and \eqref{4cr} respectively, we find that
\begin{equation}\label{5br}
\Psi \cdot p_0 =0.
\end{equation}
Equations \eqref{4er} and \eqref{4fr} reduce to
\begin{equation}\label{5cr}
w (\rho_0 +p_0) =0,
\end{equation}
and
\begin{equation}\label{5dr}
v (\rho_0 +p_0) =0.
\end{equation}
The equations \eqref{5ar}-\eqref{5dr} imply that $ \Delta =\Psi =w =v =0 $.  The only perturbing term in the matter part that remains is the azimuthal velocity $ u $ in equation \eqref{5er}.  

\section{Analysis of Results and Discussions}

In this work we have effectively combined the field equations for the perturbed background in order to evaluate the perturbation term $ h_0(t,r) $ independently, and subsequently determined $ h_1(t,r) $ by substitution. The calculations yielded a wave equation for the function $ \mathcal{Q}(t,r) $ (i.e. equation \eqref{G2} or similar equations in the different cases), which determines $ h_0(t,r) $. This wave equation not only contains the second-order derivatives of $ \mathcal{Q}(t,r) $ but also its first-order derivatives. The derivatives $ \dot{\mathcal{Q}}(t,r) $ and $ \mathcal{Q}'(t,r) $ will be responsible for the damping of the wave as it spreads out in the $(t-r)$ hypersurface. Since damping terms are absent in FLRW universe \cite{M1, M2, M3, SH1}, but are present in the LTB background (equation (50) of \cite{CL1}) in the case of axial waves, we can say that the anisotropy of the background spacetime leads to the damping of the gravitational waves in the Bianchi I spacetime considered by us.

Further, in this wave equation (in all cases), the pre-factor of $ \mathcal{Q}(t,r) $ appears as an effective potential, similar to  eqs. (4.7a) and (4.7b) of \cite{VISH}, eqs. (87) \& (91) of \cite{REZ}, or eqs. (24) and (25) of \cite{RW} for the Schwarzschild background. In this respect, the Regge-Wheeler equation exhibits all the properties of a wave equation in a scattering potential, as shown in \cite{REZ}, which dealt with gravitational waves emerging out from perturbed compact objects. Only in the general case (and the particular case denoted by \textbf{Case 1} in our paper, where $ r $ is raised to the power $ \alpha=1 $), the term $ \mathcal{Q}'(t,r) $ (the first-order $ r $-derivative of $ \mathcal{Q}(t,r) $) remains. Although the solution of the equation for $ \mathcal{R}(t,r) $ slightly changes ($ 1/r^{\alpha} $ dependence in \eqref{G6} or $1/r$ dependence in \eqref{6e1}) with the presence of $ \mathcal{Q}' $, the  $ r $-dependence explicitly inserted in the expressions for $ h_0(t,r) $ and $ h_1(t,r) $ has the same sinusoidal nature in all the four cases that we considered.

The significant difference between the solutions in the four cases lies in the $ t $-dependence. We find that the same biconfluent Heun's function appears in all the cases (see \cite{HOR, RON} for an exposition on Heun's function). But the powers of $ t $ multiplying it can assume integral or fractional values. Thus the nature of the axial perturbations $ h_0(t,r) $ and $ h_1(t,r) $ varies only with the power `$ \beta $' of $ t $ in the choice of $ \mathcal{Q}(t,r) $ (equation \eqref{G7a} in the general case and corresponding equations in the particular cases). For higher values of $l$, the  $ \mathcal{T} (t) $ solution changes due to a change in the Heun's function, leading to a change in both $ h_0(t,r) $ and $ h_1(t,r) $.

Finally, combining the solutions for $ \mathcal{T} (t) $ and $ \mathcal{R} (r) $, we arrive at the complete expressions for the axial perturbations in the form of the product of four functions, each a function of only one of the four coordinates $ t, \, r, \, \theta, \, \textrm{and} \; \phi $.
The $\phi$-dependence has been dropped by setting the projection of the angular momentum on the $ z $-axis to zero (i.e. $ m=0 $) following Regge and Wheeler \cite{RW}. The $\theta$-dependence is represented by the term $ \sin\theta (\partial_{\theta}Y)$ in \eqref{nonzero_pert}.

When the Bianchi I background is filled with matter in the form of a perfect fluid, we found that the only perturbing term in the matter part that remains is the azimuthal velocity $ u $. Thus axially perturbed Bianchi I spacetime in presence of matter will lead to perturbations only in the azimuthal velocity of the fluid without disturbing the matter field. In the RW gauge, the polar waves in FLRW universe perturb the density as well as the non-azimuthal components of the velocity of the material medium \cite{M2}, which leads to the evolution of matter inhomogeneities and anisotropies. However, for axial waves \cite{M2,SH1}, the initial data can be chosen to decouple axial perturbations from matter, causing local cosmological rotation and changes in only the azimuthal velocity. Smooth axial GWs do not interact with perfect fluids in radiative FLRW spacetimes \cite{M3}.

Another important issue is that of the equivalent GWs in the case of minimally coupled scalar fields. Cho and Speliotopoulos \cite{CHO} studied the propagation of classical GWs in Bianchi I universes and found that GWs in such spacetimes are not equivalent to two minimally coupled massless scalar fields as in the FLRW universe \cite{CHOref}. Due to its tensorial nature, the GWs are more sensitive to the background anisotropy than to the scalar field. They also found a coupling between the two polarization states of the GW which is absent in the FLRW universe. In the usual FLRW background, with a minimally coupled scalar field $ \varphi $ \cite{BARTOLO}, at the linear order, the tensor perturbations $ h_{ij} $ decouple from the scalar and vector perturbations, and satisfy the wave equation
\begin{equation}
\nabla ^2 h_{ij} -a^2 \ddot{h}_{ij} -3a\dot{a} \dot{h}_{ij} =0.
\end{equation}
The solutions for $h_{ij}$ appear in terms of the polarization tensor $ e_{ij}^{(+, \times)} $, with two polarization states, $ + $ and $ \times $. Very often a minimally coupled scalar field is considered to model the inflationary phase of the early universe, when the pressure and energy density are related as $ p \simeq -\rho $ \cite{BARTOLO}, then it appears from the right hand side of equation (20) in \cite{M2} that the axial GWs in the RW gauge will not interact with the scalar field. The scalar field $ \varphi $ with potential energy $ V (\varphi) $, drives inflation and also gives rise to gravitational perturbations. Ahmad et al studied the generation of relic gravitational waves in the paradigm of quintessential inflation \cite{ahmad} in a model-independent way, and showed that the presence of the kinetic regime after inflation results in the blue spectrum of GW background at high frequencies.

Price and Siemens \cite{PRICE} considered the generation of a stochastic background of GWs in the period following inflation in the FLRW universe. They found that the only contribution to the transverse-traceless ($ TT $)-part of the energy-momentum tensor $ T_{ij} $ comes from the anisotropic stress $ \pi_{ij}^{TT} $, which becomes the source term for GWs in this case. It has also emerged that GWs (tensor perturbations) in FLRW universe cannot be sourced by background fields and their perturbations \cite{MUKHANOV, ZHANG}. Thus it can be said that GWs in the $ TT $ gauge are not affected by the presence or absence of scalar fields or matter content unless there is anisotropic stress.

\section{Summary and Conclusions}
From the above analysis, it is evident that the Regge-Wheeler gauge is applicable to the Bianchi I metric for small values of the polar angle $ \theta $.
Previous studies on the propagation of GWs in Bianchi I universe have been undertaken in the Cartesian coordinate system. Different gauges have been employed and different formalism for the decomposition of perturbations have been proposed in a number of papers. However, the Regge-Wheeler gauge have not been applied to study gravitational perturbations in the Bianchi I universe.

We have solved for the axial perturbations $ h_0(t,r) $ and $ h_1(t,r) $ in Regge-Wheeler gauge for the Bianchi I spacetime. We have expressed the line element in spherical polar coordinates. This is required for applying the Regge-Wheeler gauge. The axial perturbation terms $ h_0(t,r) $ and $ h_1(t,r) $ in this gauge are obtained explicitly as a product of four factors, each a function of one coordinate only. The perturbed field equations led to a wave equation with damping, resulting from the anisotropy of the background. Similar wave equations have been obtained in \cite{M1, M2, M3, SH1, CL1} for the FLRW spacetime, or in \cite{REZ} for a Minkowski background. While solving the linearized Einstein equations for the perturbed metric, the remaining $ t $ and $ r $-dependence of $ h_{\mu\nu} $ separated out as a product. We find that the biconfluent Huen's function appears in the solution of the temporal part. The corresponding solution in the case of the Schwarzschild metric emerged in terms of the confluent Heun's function \cite{FIZ}.
In order to study the axial modes coming from the expansion of both vector and tensor functions, we have considered $l\geq 2$ for our analysis. We found that a variation in $l$ will produce a change in both  $ h_0(t,r) $ and $ h_1(t,r) $.

When the Bianchi I background is filled with perfect fluid, we found that the axially perturbed spacetime will lead to perturbations only in the azimuthal velocity of the fluid leaving the matter field undisturbed. A comparison of the equivalent GWs in case of minimally coupled scalar field for the FLRW universe with that in the Bianchi I universe reveals that GWs in the FLRW universe are equivalent to two minimally coupled massless scalar fields whereas in the Bianchi I universe the GWs are more sensitive to the background anisotropy than to the scalar field and the two polarization states are coupled. It appears that the GWs are not affected by the presence or absence of scalar fields or matter content unless there is anisotropic stress.

We intend to extend our work by considering a matter distribution in the case of polar gravitational waves in the Bianchi I universe.

\section{Acknowledgements}
The authors are thankful to the anonymous reviewers for the useful comments to improve the quality of the paper. SG thanks IUCAA, India for an associateship, and CSIR, Government of India for financial support. SD acknowledges the financial support from INSPIRE (AORC), DST, Govt. of India (IF180008).


\begin{thebibliography}{}

\bibitem{EINST1} A. Einstein, Sitzungsber. Preuss. Akad. Wiss. Berlin (Math. Phys.) 1916, 688 (1916).
\bibitem{EINST2} A. Einstein, Sitzungsber. Preuss. Akad. Wiss. Berlin (Math. Phys.) 1918, 154 (1918).
\bibitem{Rosen} N. Rosen, Phys. Z. Sovjet Union \textbf{12}, 366 (1937).
\bibitem{ER} A. Einstein and N. Rosen, Journ. of Franklin Institute \textbf{223}, 43 (1937).
\bibitem{Yvonne} Y. Choquet-Bruhat, Acta Math. \textbf{88}, 141 (1952).

\bibitem{Bondi} H. Bondi, Nature \textbf{179}, 1072 (1957).

\bibitem{Pirani} F. A. E. Pirani, Phys. Rev. \textbf{105}, 1089 (1957).

\bibitem{BPR} H. Bondi, F. A. E. Pirani, and I. Robinson, Proc. R. Soc. London Ser. A \textbf{251}, 519 (1959).

\bibitem{RT1} I. Robinson and A. Trautman, Phys. Rev. Lett. \textbf{4}, 431 (1960).
\bibitem{RT2} I. Robinson and A. Trautman, Proc. Roy. Soc. London Ser. A \textbf{265}, 463 (1962).

\bibitem{ST} M. Sasaki and H. Tagoshi, Living Rev. Relativity \textbf{6}, 6 (2004), [Online article]: cited on http://www.
livingreviews.org/lrr-2003-6.


\bibitem{RW} T. Regge  and J. A. Wheeler, Phys. Rev. \textbf{108}, 1063 (1957).
\bibitem{ZER} F. J. Zerilli, Phys. Rev. Lett. \textbf{24}, 737 (1970). 
\bibitem{VISH} C. V. Vishveshwara, Phys. Rev. D \textbf{1}, 2870 (1970).

\bibitem{Chandra} S. Chandrasekhar, The Mathematical Theory of Black Holes (Clarendon Press, Oxford, 1983).

\bibitem{MP} K. Martel and E. Poisson, Phys. Rev.  D \textbf{71}, 104003 (2005).

\bibitem{VIA} S. Viaggiu, Class. Quantum Grav. \textbf{34}, 035018 (2017).
\bibitem{MON} V. Moncrief, Annals of Physics \textbf{88}, 323 (1974).
\bibitem{FIZ} P. P. Fiziev,  Class. Quantum Grav. \textbf{23}, 2447 (2006); arXiv:gr-qc/0509123.
\bibitem{M1} E. Malec  and G. Wylezek, Class. Quantum Grav. \textbf{22}, 3549 (2005).
\bibitem{M2} W. Kulczycki  and  E. Malec, Class. Quantum Grav. \textbf{34}, 135014 (2017).
\bibitem{M3} W. Kulczycki  and  E. Malec, Phys. Rev. D \textbf{96}, 063523 (2017).
\bibitem{SH1} M. Sharif and A. Siddiqa, Eur. Phys. J. C. \textbf{78}, 721 (2018); arXiv:1809.03307 [gr-qc].
\bibitem{SH2} M. Sharif and A. Siddiqa, Gen. Relativ. Gravit. \textbf{51}, 74 (2019); arXiv:1907.05714 [gr-qc].


\bibitem{GMG} C. Gundlach and J. M. Martin-Garcia, Phys. Rev. D \textbf{61}, 08402 (2000).
\bibitem{GS} U.H. Gerlach, and U. K. Sengupta, Phys. Rev. D \textbf{19}, 2268 (1979); ibid \textbf{22}, 1300 (1980).
\bibitem{CL1} C. Clarkson, T. Clifton  and S. February,  J. Cosmol. Astropart. Phys. 06(2009)25.
\bibitem{CL2} S. February, J. Larena, C. Clarkson, D. Pollney, Class. Quantum Grav. \textbf{31}, 175008 (2014).
\bibitem{BARDEEN} J. M. Bardeen, Phys. Rev. D \textbf{22}, 1882 (1980).
\bibitem{STEWART} J. M. Stewart, Class. Quant. Grav. \textbf{7}, 1169 (1990).
\bibitem{MUKHANOV} V. F. Mukhanov, H. A. Feldman and R. H. Brandenberger, Phys. Rept. \textbf{215}, 203 (1992).
\bibitem{MIEDEMA96} P. G. Miedema and W. A. van Leeuwen, Phys. Rev. D \textbf{54}, 7227 (1996).
\bibitem{BR1} Z. Keresztes, M. Forsberg, M. Bradley, P. K. S. Dunsby, and L. A. Gergely, J. Cosmol. Astropart. Phys. 11(2015)042.
\bibitem{BR2} M. Bradley, M. Forsberg, and Z. Keresztes, Universe \textbf{3}, 69 (2017).

\bibitem{EM} G.F.R. Ellis and M.A.H. MacCallum, Commun. Math. Phys. \textbf{12}, 108 .(1969)
\bibitem{Bianchi} L. Bianchi, Soc. Ital. Sci. Mem. Mat. \textbf{11}, 267 (1898).

\bibitem{HU} B. L. Hu, Phys. Rev. D \textbf{18}, 968 (1978).
\bibitem{MIEDEMA93} P. G. Miedema and W. A. van Leeuwen, Phys. Rev. D \textbf{47}, 3151 (1993).
\bibitem{CHO} H. T. Cho and A. D. Speliotopoulos, Phys. Rev. D \textbf{52}, 5445 (1995); arXiv:gr-qc/9504046.
\bibitem{CHOref} L. H. Ford and L. Parker, Phys. Rev. D \textbf{16}, 1601 (1977).
\bibitem{ADAMS1} P. J. Adams, R. W. Hellings, R. L. Zimmerman, H. Farhoosh, D. I. Levine and S. Zeldich, Astrophys. J. \textbf{253}, 1 (1982).
\bibitem{ADAMS2} P. J. Adams, R. W. Hellings and R. L. Zimmerman, Astrophys. J. \textbf{288}, 14 (1985).

\bibitem{PER} T. S. Pereira, C. Pitrou and J.-P. Uzan, JCAP 0709(2007)006;  arXiv:0707.0736 [astro-ph].
\bibitem{PELO} A. E. G\"umr\"uk\c c\"u\u glu, C. R. Contaldi and M. Peloso, JCAP 0711 (2007) 005; arXiv:0707.4179 [astro-ph].


\bibitem{Jacobs} K. C. Jacobs, Ph.D. thesis, California Institute of Technology, Pasadena, 1969.

\bibitem{SHAMIR1006} M. F. Shamir, Astrophys. Space Sci. \textbf{330}, 183 (2010);
arXiv:1006.4249v1 [gr-qc].

\bibitem{Gron} {\O}. Gr{\o}n, Phys. Rev. D \textbf{32}, 2522 (1985).
\bibitem{ans1} S. R. Roy, S. Narain, J. P. Singh, Aust. J. Phys. \textbf{38}, 239 (1985).
\bibitem{ans2} S. R. Roy, S. K. Banerjee, Class. Quantum Gravity \textbf{11}, 1943 (1995).
\bibitem{ans3} P. S. Baghel and J. P. Singh, Int. J. Theor. Phys. \textbf{51}, 3664 (2012).
\bibitem{ans4} R. Bali, R. Banerjee, S. K. Banerjee, Astrophys. Space Sci. \textbf{317}, 21 (2008).


\bibitem{CGW} C. B. Collins, E. N. Glass and D. A. Wilkinson, Gen. Relativ. Gravit. \textbf{12}, 805 (1980).

\bibitem{Maple} These calculations were done using Maple software.

\bibitem{HOR} M. Horta\c csu, Proceedings of the 13th Regional Conference on Mathematical Physics, Antalya, Turkey, October 27–31, 2010, (Eds. U. Camc and I. Semiz), World Scientific, Singapore, 2013, p. 23; arXiv: 1101.0471.
\bibitem{RON} A. Ronveaux (Ed.), Heun’s Differential Equations, Oxford University Press, Oxford 1995.

\bibitem{BARTOLO} M. C. Guzzetti, N. Bartolo, M. Liguori and S. Matarrese, Rivista del Nuovo Cimento \textbf{39}, 399 (2016).

\bibitem{ahmad} S. Ahmad, R. Myrzakulov, and M. Sami, Phys. Rev. D \textbf{96}, 063515 (2017).

\bibitem{PRICE} L. R. Price and X. Siemens, Phys. Rev. D \textbf{78}, 063541 (2008).

\bibitem{ZHANG} N. Zhang and Y. E. Cheung, Eur. Phys. J. C \textbf{80}, 100 (2020).

\bibitem{REZ} L. Rezzolla, Lecture notes : Gravitational Waves from Perturbed Black Holes and Relativistic Stars, Summer School on Astroparticle Physics and Cosmology, Trieste, 17 June - 5 July, 2002.

\end{thebibliography}
\end{document}